# An SDR Implementation of a Visible Light Communication System Based on the IEEE 802.15.7 Standard

Jorge Baranda*, Pol Henarejos*, Ciprian G. Gavrincea*
*Centre Tecnològic de Telecomunicacions de Catalunya (CTTC), Castelldefels, Barcelona (Spain)
Email: {jorge.baranda, pol.henarejos, ciprian.gavrincea}@cttc.es

*Abstract*—The aim of this paper is to present an implementation of a functional IEEE 802.15.7 real-time testbed based on the Software Defined Radio (SDR) concept. This implementation is built with low cost commercial off-the-shelf (COTS) analog devices and the use of Universal Software Radio Peripheral version 2 (USRP2) equipment combined with a generic object-oriented framework that takes advantage of several open source software libraries. The prototype is validated in a controlled laboratory environment with an over-the-air measurement campaign.

## I. Introduction

Visible Light Communications (VLC) is an emerging research field of free space optic (FSO) communications. The evolution of high-power white light emitting diodes (LEDs) within the last decades has led to the development of low cost lighting devices with better performance in terms of both energy efficiency and life expectancy. LEDs are expected to replace incandescent light bulbs and fluorescent lamps in the next generation illumination systems. Nonetheless, these devices can be used for applications beyond the illumination purpose [1]. The use of a LED to transmit data while serving as an illumination source has captured the interest of the research community as well as global standardization efforts [2].

The use of VLC technology presents many attractive advantages in terms of available bandwidth, non-interference with radio bands (critical especially due to the scarcity in some frequency bands such as ISM), potential spatial reuse, inherent protection against eavesdropping and, all of this with power efficient devices and without high deployment costs. In the last years, the focus of the research community has been centered in improving the data rates achievable by VLC systems, obtaining performances of 1.25 Gb/s [3]. Nevertheless, although there are some standards already published for VLC communications ([4], [5]), proof-of-concept implementation examples based on these standards are scarce.

In this paper, we present, to the best of our knowledge, the first implementation of a real-time VLC system prototype based on the IEEE 802.15.7 specification [5] using low cost COTS analog devices. For this development, we have followed an SDR based design approach, similar to the one presented in [6]. The SDR based approach brings a lot of advantages to develop communication system testbeds, regarding a pure hardware (HW) based implementation. Among them, we would like to highlight: flexibility, modularity, reduced financial costs and shorter development time. These advantages are due to the fact that, in an SDR-based implementation, hardware problems are turned into software problems, allowing the use of novel algorithms almost immediately, accelerating the transition from simulation to demonstration. From the variety of available commercial SDR platforms [7], we find attractive the equipment provided by Ettus Research [8], specially the USRP2 and its predecessors, because of its superior trade-off between price and performance and, the support provided by a large user community. Finally, this paper unveils the performance of the developed VLC system with an over-the-air measurement campaign.

## II. IEEE 802.15.7 PHY layer brief Overview

The IEEE 802.15.7 standard [5] defines the physical (PHY) and the medium access control (MAC) layer for short-range optical wireless communications using visible light for indoor and outdoor applications. One of its main distinguishing points compared to other specifications, such as the Japan Electronics and Information Technology Industries (JEITA) specifications [4], is the support for dimming and flickering mitigation techniques [9]. These aspects are very important due to the negative physiological effects that flickering may have in humans and because dimming allows power savings and energy efficiency.

The standard presents three different PHY layers which are grouped by their data rates. The operation range of PHY I is from 11.67 kb/s to 266.6 kb/s, PHY II data rates are comprised between 1.25 Mb/s and 96 Mb/s, while PHY III operates from 12 Mb/s to 96 Mb/s. The modulation formats are different among PHY I/II and PHY III. While PHY I and PHY II use on-off keying (OOK) and variable pulse-position modulation (VPPM), PHY III uses a particular modulation format called color shift keying (CSK), in which information is transmitted by means of multiple optical sources.

Each PHY mode includes run length limiting (RLL) codes to achieve direct current (DC) balance, avoiding data induced flickering, and forward error correction (FEC) schemes to improve the link reliability for the envisaged environments. PHY I is optimized for low data rate, transmissions over long distances, e.g., outdoor applications such as traffic and vehicle lights; and PHY II/III are designed for high data rate indoor point-to-point applications. FEC schemes proposed by the standard are a combination of Reed-Solomon (RS) and convolutional codes (CC). These schemes work reasonably well in the presence of hard-decisions performed by the clock data recovery circuits (CDR) and make the interface with the proposed RLL codes (Manchester, 4B6B and 8B10B) easier.

### III. DEVELOPMENT METHODOLOGY

The development of this prototype started with the implementation of a reference encoder-decoder model in MATLAB according to the specifications of [5] for PHY I/II operating modes. Regarding the MAC layer, only a set of features of the header specification are included to simplify the communication link. The construction of the model was useful for a better understanding of the specification and constituted an important debugging tool which helped in the posterior implementation under the FlexiCom object-oriented framework.

FlexiCom framework is born as an evolution of the uPHYLA framework [10]. As uPHYLA, FlexiCom framework is intended as a tool to implement in software the physical layer of any communication system. Hence, a practical demonstration of novel signal processing and communication algorithms with real digitized signals can be performed within a short development time. Nevertheless, the new created architecture for the FlexiCom framework provides enhanced graphical capabilities and simplicity to users and developers, as depicted in Fig. 1.

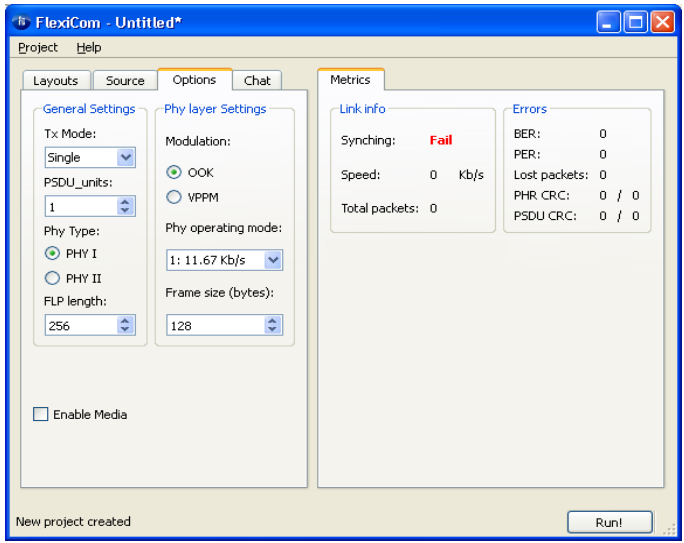

Fig. 1. Graphical Interface used to configure the VLC communication layout.

FlexiCom platform uses Nokia's Qt library [11] to offer the possibility of choosing among a set of communication system layouts. Each layout presents a customized interface for system configuration and performance metric visualization. Moreover, this framework runs over most operating systems (OS) such as Windows, Linux or Mac OSX. Once the system is configured, FlexiCom uses the GNU Radio library [12] to construct the communication system and to distribute the samples among the involved blocks in a continuous flow.

In the second development phase, the MATLAB code is ported to the FlexiCom framework. Due to the stream-based flow of samples in the GNU Radio environment, the effort is focused on redesigning the encoder/decoder chains, identifying the minimal processing units and introducing low-complexity algorithms which improve the system latency.

Finally, we would like to highlight the importance of the testing phase to verify the successive development steps. In the MATLAB programming step, the exhaustive test of all possible combinations allows the functional verification of the encoder and the decoder implementation. Moreover, the information provided by this initial testing phase helps to accelerate the multi-level testing phase (basic block unit testing, module testing and system testing) of the implementation under the FlexiCom framework by means of the CppUnit testing framework [13].

### IV. SYSTEM MODEL

The current prototype implementation supports the modulation and demodulation of all the operating and data modes defined for PHY I at [5], both for OOK and VPPM schemes. These modes are included in Table I. Based on the SDR approach used in this implementation, the prototype is divided into two main subsystems: hardware and software subsystem, as shown in Fig. 2. The hardware subsystem is constituted by the elements labeled as LED, Photodiode, USRP TX and USRP RX, while the software subsystem is composed by the elements labeled as FlexiCom TX and FlexiCom RX.

TABLE I
PHY I OPERATING MODES

| Modulation | RLL Code | Optical clock rate | FEC Outer Code (RS) | FEC Inner Code (CC) | Data Rate | Operating mode |
|---|---|---|---|---|---|---|
| OOK | Manchester | 200 KHz | (15,7) | 1/4 | 11.67 kb/s | 1 |
| | | | (15,11) | 1/3 | 24.44 kb/s | 2 |
| | | | (15,11) | 2/3 | 48.89 kb/s | 3 |
| | | | (15,11) | none | 73.3 kb/s | 4 |
| | | | none | none | 100 kb/s | 5 |
| VPPM | 4B6B | 400 KHz | (15,2) | none | 35.56 kb/s | 1 |
| | | | (15,4) | none | 71.11 kb/s | 2 |
| | | | (15,7) | none | 124.4 kb/s | 3 |
| | | | none | none | 266.6 kb/s | 4 |

#### A. Hardware Subsystem

The analog transmitter front-end is constituted by the USRP2 platform equipped with the LFTX daughterboard, an amplification stage, the LED driver circuit and a commercial high power white LED as the light emitting source.

The software subsystem generates the stream of bits which are digital to analog (DAC) converted in the USRP2. The modulated signal provided by the SDR platform is amplified to be adjusted to the levels required to control the LED driver circuit. This circuit is composed of a power MOSFET transistor (STD12NF06L) which acts in switching mode, providing the needed current to the LED fixture. The light source is a commercial phosphorescent white LED module (OSTAR® LE CW E2B), consisting of a blue LED covered by a yellow phosphor layer designed for indoor lighting applications. The light source is equipped with a reflector that reduces the viewing angle of the LED and concentrates the light beam. With the current reflector, the viewing angle is reduced to 30° (half optical transmitted power is achieved at 15° from the axis of the LED device).

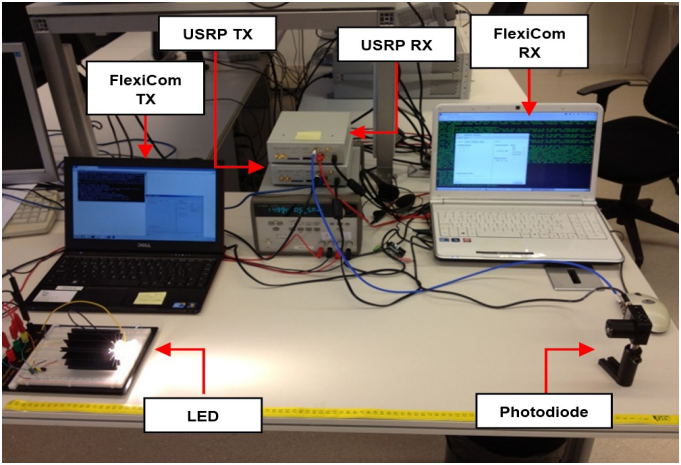

Fig. 2. IEEE 802.15.7 based VLC Prototype.

The analog receiver front-end is constituted by the USRP2 device equipped with the LFRX daughterboard and a commercial photodetector (THORLABS® PDA36A) equipped with a spherical concentrating lens. The detector consists of a PIN silicon photodiode with an active area of 13 mm$^2$ and an optical sensitivity of 0.2-0.4 A/W in the visible range. The PDA36A device is equipped with a transimpedance amplifier (TIA) chain circuit with manual adjustable gain, ranging between 0 and 70 dB. This setting impacts in the receiver bandwidth. But this is not the only bandwidth limiting factor of this setup. Beside the limitations of the LED fabrication process [14], the high amount of current that is required to operate ($\sim$ 700 mA) conditions the switching speed possibilities of the led driver circuit.

*B. Software Subsystem*

The software subsystem is constituted by two general purpose processors (GPPs), one for the modulation and one for the demodulation of the signals. An IEEE 802.15.7 frame is composed of three elements: the synchronization header (SHR), the physical header (PHR) and the physical service data unit (PSDU). In order to maintain the level of luminance and to avoid inter-frame flickering, an in-band idle pattern is inserted between frames [9]. The SHR contains the fast locking pattern (FLP) and the topology dependent pattern (TDP), used to lock the CDR circuit and perform the synchronization with the incoming message. The PHR contains frame information related to the length of the data unit, the used modulation and FEC scheme. The PSDU is the data unit, and also contains the MAC Header (MHR) in which parameters such as the sequence length and control format fields are included. As the current implementation features a broadcasting system, it modulates/demodulates the MHR headers according to the format specified in [5] for data frames. The building blocks of the physical layer transmitter and receiver are depicted in Fig. 3 and Fig. 4, respectively.

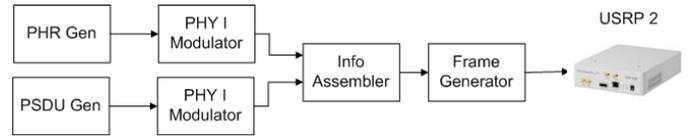

Fig. 3. Structure of the VLC IEEE 802.15.7 software transmitter.

At the transmitter, the PHR and the PSDU units are generated independently according to the system configuration. Each unit passes through a PHY I modulator block, where the corresponding FEC scheme and RLL coding are applied (see Table I). After the modulation, both units are assembled into one stream of data and the TDP pattern is inserted. In the last block, before the USRP2 device, the FLP pattern and the idle pattern are attached to form the entire frame.

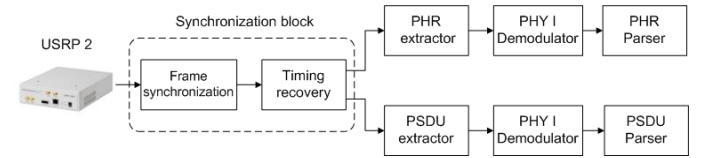

Fig. 4. Structure of the VLC IEEE 802.15.7 software receiver.

At the receiver, frame synchronization and timing recovery are performed at the first block. Frame recovery is achieved through the exploitation of repeated patterns at the SHR header. Timing recovery is the procedure to determine the optimum sampling instant required to decide the value of the incoming bits. Due to the limitations of the USRP2 hardware in terms of clock reconfigurability, timing recovery is performed using a non-data aided detector based on the maximum likelihood (ML) algorithm [15]. According to this detector, the optimum sample corresponds to the one that maximizes the energy of the received oversampled sequence of samples. As this prototype works with a binary modulation, many transitions are produced within a frame. The oversampling allows a better reconstruction of the waveform, so the transitions can be easily detected through the maximum energy criterion.

Both algorithms have a limited complexity, allowing a feasible implementation in terms of system latency. Moreover, the support of Vector-Optimized Library of Kernels (VOLK) inside the GNU Radio library allows the use of Streaming SIMD Extensions (SSE), which boosts the execution of the synchronization block by a factor of four. Once the incoming flow of bits is synchronized, the PHR and the PSDU units are extracted and demodulated according to the system configuration. Finally the bits are parsed in order to extract the information corresponding to the different units and the validity of the received frame is checked. A frame is declared invalid if there is an error in the cyclic redundancy check (CRC) field, or the frame sequence number does not correspond to the previous frame.

Fig. 5 shows the distribution of the relative computational load among the receiving processing blocks obtained with the OProfile tool [16] when using Operation mode 1 and OOK modulation, which is the more demanding operating mode in terms of computer processing. The execution of the receiving chain requires only 25% of the total computer processing power. The synchronization block represents 55% of the processing, while the demodulation (Manchester + CC + RS decoding + Deinterleaver operations) accounts for the 35%. The remaining 10% is required to extract the PHR and the PSDU bit streams and to parse the information included within these units.

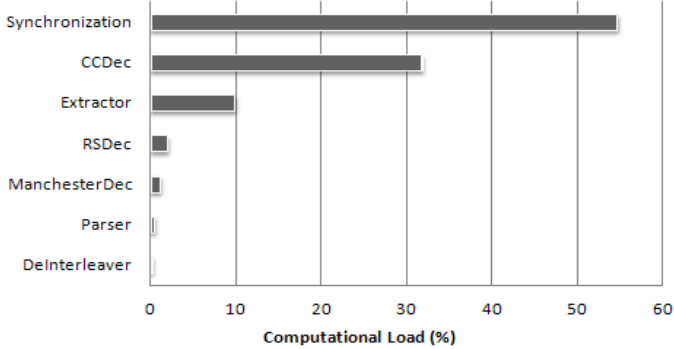

Fig. 5. Distribution of the computational load among the different receiving processing blocks.

## V. RESULTS

Once all the system units are tested and validated individually, the last phase of the development is dedicated to characterize the performance of the complete implementation (baseband modulation/demodulation, signal conditioning (DAC/ADC) and analog optical front-ends) using over-the-air measurements in a controlled environment. The transmitter code runs in a PC laptop with 2 GB of RAM and an Intel Core 2 Duo I3 processor clocked at 1.5 GHz. The receiver code runs in a PC laptop with 4 GB of RAM and an Intel Quad-core I5 processor clocked at 2 GHz. Additional details of the test-system are summarized in Table II.

TABLE II
TESTBED SYSTEM DETAILS

| Feature | Transmitter | Receiver |
|---|---|---|
| Operating System | Windows 7 H.E. | Ubuntu 12.04 |
| Platform | FlexiCom | FlexiCom |
| QT Version | 4.8 | 4.8 |
| GNU Radio Version | 3.6.2 | 3.6.2 |
| USRP2 HW revision | 4 | 4 |
| USRP2 Firmware version | 3.0.4 | 3.0.4 |
| Daugtherboard model | LFTX | LFRX |

In the first experiment, measurements are taken in a medium-sized room without windows where the transmitter and the receiver are horizontally aligned and separated in order to determine the system bit-error rate (BER) performance versus the distance. In this setup, the transmitter is moved in 50 cm steps and, at least, 50000 frames of 128 bits of data are transmitted to measure the system performance. The amplification of the TIA chain at the receiver is set to 0 dB. The BER performance versus distance for each operating mode of the PHY I specification using OOK modulation is presented in Fig. 6. The points presenting a BER value of $10^{-6}$ means that no error was found during the test. $10^{-6}$ constitutes the floor of our measurement because the amount of transmitted bits in each repetition is in the order of $10^6$.

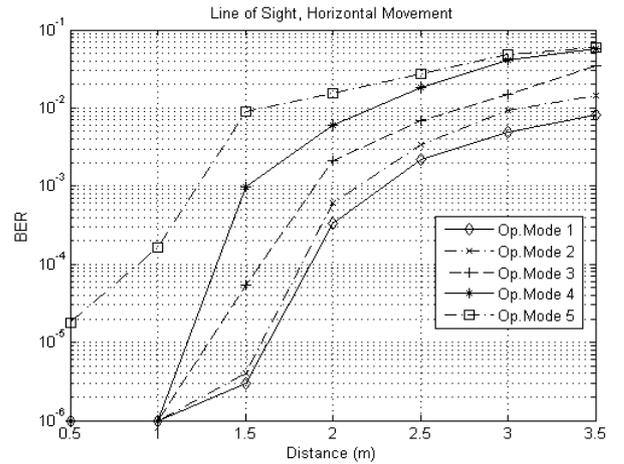

Fig. 6. Performance of the VLC IEEE 802.15.7 prototype for different operating modes of PHY I using OOK modulation.

Prior to analyze the performance of the system, a comment about the tested distances is worthy to be done. According to Section II, PHY I is thought to be used for outdoor applications. In such scenarios, the light emitter is constituted by an array of LED devices, which are capable of providing much more luminous flux than the current setup herein presented. One must remember that the current setup uses a single luminary with a maximum output of 450 lumens, according to its datasheet, which is suitable for indoor illumination systems.

As expected, the system performance degrades with the distance as a consequence of a lower SNR value at the receiver site. Operation mode 1 and 2, which present the strongest channel coding schemes among the considered operating modes (see Table I), obtain the best performance. Nonetheless, as the distance between the transmitter and the receiver increases, the impact of channel coding in less robust schemes disappears and the BER performance becomes almost equivalent to the scheme without channel coding scheme (Op. mode 5), impacting in the achieved effective data rate. We define the effective data rate as the total amount of correct information bits received without taking into account the overhead introduced by the different headers (SHR, PHR and MHR) and the idle-pattern. Table III shows the measured effective data rate for each considered operating mode using OOK modulation at a distance of two meters.

TABLE III
MEASURED EFFECTIVE DATA RATE AT 2 METERS DISTANCE

| Operating Mode | Effective data rate |
|---|---|
| Op. Mode 1 | 7.9 kb/s |
| Op. Mode 2 | 18.01 kb/s |
| Op. Mode 3 | 24.63 kb/s |
| Op. Mode 4 | 32.38 kb/s |
| Op. Mode 5 | 35.22 kb/s |

In the second experiment, we tested the end-to-end performance of our prototype when transmitting and receiving a video coded in MPEG-TS (Transport Stream) format in a medium-sized office room with standard illumination conditions. The transmitter extracts the information from an MPEG-TS stream coded at an MPEG2 video compression bit rate of 25 kb/s and audio rate of 16 kb/s, and modulates it according to the frame structure explained in Section IV-B using the coding scheme corresponding to 73.3 kb/s of data rate (see Table I). At the receiver side, the data payload is demodulated and sent through a user data protocol (UDP) socket connection to the VideoLAN streaming software [17]. Additionally, the system clock data rate is set to 500 KHz, 2.5 times faster than the IEEE 802.15.7 specification of PHY I OOK modulation in order to cope with the extra signaling of the MPEG stream and to achieve on-line displaying. The current prototype implementation supports the on-line streaming at a distance of around 1.5 m with an effective measured data rate of about 107 kb/s with a packet error rate (PER) of 1e-3. In this setup, the amplification of the TIA chain at the receiver is set to 20 dB. Fig. 7 presents one captured image from the received video signal, together with some metrics captured with the FlexiCom framework: effective data rate, number of total/lost packets, packet error rate (PER) and quantity of correct/incorrect decoded packets (PHR and PSDU units).

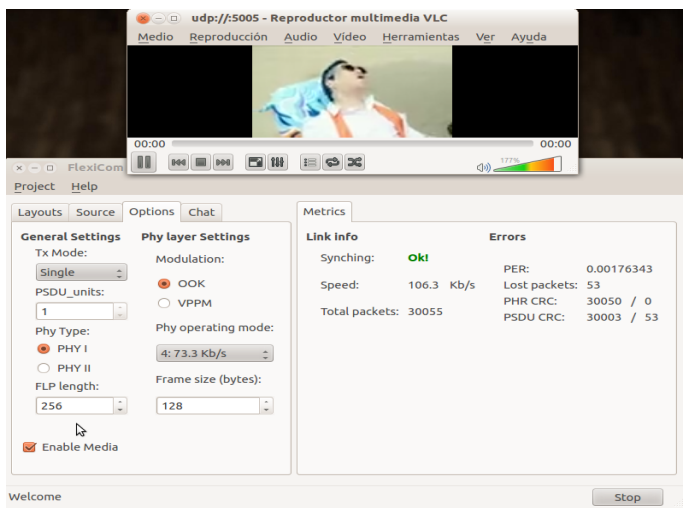

Fig. 7. Received video capture, for a 1.5m link (from "Gangnam Style", PSY, 2012).

## VI. Conclusions and future work

We have presented a functional system prototype implementing all the operating modes of the PHY I specification of the IEEE 802.15.7 standard for VLC. This development is based on the SDR concept, providing the prototype with enough flexibility and modularity to enhance its capabilities without requiring long development time and high financial costs. The prototype is completed with low cost COTS analog devices to perform the transmission and reception of the light signal. The IEEE 802.15.7 standard serves as a promising platform to start developing and introducing VLC solutions into the market for low-medium data rate point-to-point and broadcast systems in the field of intelligent transportation systems (ITS) or smart indoor locating (SIL).

Future system development will be focused on two aspects: a) the integration of dimming functionalities into the communication framework when using OOK modulation by inserting compensation symbols within the frame without reducing the range of the communication system; and b) further research into the analog subsystem to improve the system bandwidth and to implement the operating modes of PHY II. From the software point of view, the MATLAB simulator and the FlexiCom modules are easily customizable.

## Acknowledgment

The authors would like to thank Miquel Payaró, Christian Ibars and the Access Technologies area members of CTTC for their support for the implementation of the VLC system prototype.